\documentclass{article}

\usepackage{spconf,amsmath,graphicx}
\usepackage[utf8]{inputenc}
\usepackage{amsmath}
\usepackage{algorithm}
\usepackage{algorithmic}
\usepackage{commath}

\title{\bf Parallel optimization of fiber bundle segmentation for massive tractography datasets}

\name{\begin{tabular}[t]{c@{\extracolsep{1em}}c@{\extracolsep{1em}}c@{\extracolsep{1em}}c}
  \multicolumn{4}{c}{Andrea Vázquez$^{1}$, Narciso López-López$^{1,2}$, Nicole Labra$^{3}$, Miguel Figueroa$^{1}$}\\
  \multicolumn{4}{c}{Cyril Poupon$^{3}$, Jean-Fran\c{c}ois Mangin$^{3}$, Cecilia Hernández$^{1}$ and Pamela Guevara$^{1}$}
   \thanks{This work has received funding from CONICYT FONDECYT 1161427, CONICYT PFCHA/ DOCTORADO NACIONAL/2016-21160342, CONICYT PIA/Anillo de Investigación en Ciencia y Tecnología ACT172121, CONICYT BASAL FB0008, CONICYT Basal FB0001, and from the European Union’s Horizon 2020 research and innovation programme under the Marie Sklodowska-Curie grant agreement No 690941.}
\end{tabular}}

\address{$^{1}$ Faculty of Engineering, Universidad de Concepción, Concepción, Chile \\
	     $^{2}$Dept.of Computer Science, Universidade da Coruña, A Coruña, Spain \\
	     $^{3}$I2BM, Neurospin, CEA, Gif-sur-Yvette, France}

\begin{document}

\maketitle

\begin{abstract}
We present an optimized algorithm that performs automatic classification of white matter fibers based on a multi-subject bundle atlas. We implemented a parallel algorithm that improves upon its previous version in both execution time and memory usage. 
Our new version uses the local memory of each processor, which leads to a reduction in execution time. Hence, it allows the analysis of bigger subject and/or atlas datasets.
As a result, the segmentation of a subject of 4,145,000 fibers is reduced from about 14 minutes in the previous version to about 6 minutes, yielding an acceleration of 2.34. In addition, the new algorithm reduces the memory consumption of the previous version by a factor of 0.79.

\end{abstract}


\section{Introduction}
\label{sec:intro}

The study of white matter (WM) structure is a constantly-growing research area, as it aims to understand brain connectivity and its relation to function. Diffusion-weighted Magnetic Resonance Imaging (dMRI) is the preferred technique for the study of anatomical connectivity, through the measurement of water molecule movement in a non-invasive way \cite{LeBihan01}. Diffusion tractography estimates the trajectory of the main WM pathways in the brain, leading to datasets that contain a large number of 3D trajectories, called fibers or streamlines. Contrary to voxel-based approaches, tract-specific methods extract bundles from tractography datasets to study the structural connectivity in healthy and pathological brains, for example, for the study of psychiatric disorders \cite{Sarrazin14}.

Bundle segmentation methods, which label the main brain fiber bundles, are implemented following two main strategies.
First, a number of anatomical regions of interest (ROIs) can be used to identify the fibers that connect two or more regions in the brain. This process can be done automatically \cite{ZhangY10, Wassermann2016} or manually by an expert \cite{Catani02, catani2012short}.
The second strategy uses a fiber distance metric to match similar fibers, taking into account their shape and relative position \cite{O'Donnell07, Wassermann10}. These methods add anatomical information to identify fiber bundles with anatomical significance. 
This information, along with fiber shape and position for multiple subjects, can be embedded on a multi-subject atlas \cite{Guevara12}. Segmentation based on this kind of atlas can deal with the variability that exists between the subjects, and achieves good results in fiber classification.

The segmentation algorithm proposed in \cite{Guevara12} was implemented in the C programming language, and exploits thread-level parallelism on multiples CPUs to deal with massive tractography datasets and achieve short execution time \cite{labra2017fast}. Our goal in this work was to improve the algorithm   and its implementation, aiming to reduce the use of computing resources, mainly CPU and memory. We wrote the new implementation using the C++ programming language, which improves upon the modularity and extensibility of the previous version. Optimizations were focused on reducing the memory used for temporary results, improving task parallelization and the fast bundle-discarding algorithm, and performing the classification of fibers using only the closest bundle.

These improvements allow us to perform segmentation of very large datasets on a desktop computer. Furthermore, the new algorithm is more scalable, thus enabling segmentation on more massive tractography (and atlas) datasets, in particular, those obtained from superficial white matter (SWM) fibers with probabilistic tractography, which can reach several millions of fibers.


\section{Materials and Methods}
\label{sec:matmet}

\subsection{Database and tractography datasets} 
We used healthy subjects from the ARCHI HARDI database \cite{Schmitt12}, containing T1, HARDI and fMRI images, acquired with special acquisition sequences on a 3T MRI scanner (Siemens, Erlangen). The MRI protocol included a HARDI SS-EPI single-shell dataset along 60 optimized DW directions, b=1500 $s/mm^2$ (70 slices; matrix=128x128; voxel size=1.71875x1.71875x1.7 mm).
The data were pre-processed using BrainVISA/Connectomist-2.0 software. Analytical Q-ball model was computed to obtain ODF fields in each voxel and streamline deterministic tractography was performed on the entire T1-based brain mask, with a forward step of 0.2 mm and a maximum curvature angle of 30$^\circ$.



\subsection{Multi-subject atlases}
SWM multi-subject atlases where constructed based on intra-subject cluster centroids \cite{Guevara11} from a population of subjects. The LNAO-SWM79 atlas was obtained from a HARDI database of 79 subjects \cite{Schmitt12}, based on a cortical parcellation to extract the fibers that connect two regions, followed by an intra- and inter-subject fiber clustering (100 bundles, 7.753 centroids) \cite{guevara2017reproducibility}. 
The other atlas was obtained using the same database, but applying a whole-brain clustering strategy and non-linear registration (62 bundles, 44.345 centroids) \cite{roman2017clustering}. The centroids in both atlases have 21 equidistant points.

\begin{figure}[t!]
	
	\begin{minipage}[b]{1.0\linewidth}
		\centerline{\includegraphics[width=8.5cm]{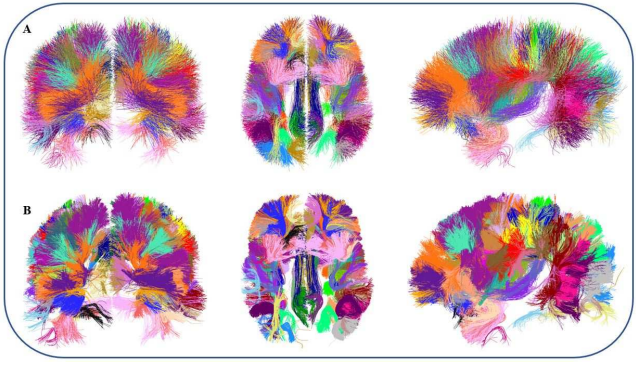}}
		\caption{SWM bundle atlas and a subject segmented with the new algorithm (coronal, axial and sagittal views). A. SWM bundle atlas (LNAO-SWM79), composed  of  50  bundles per hemisphere. B. Segmented subject of 955K fibers.}
		\label{fig:atlasSubject}
	\end{minipage}
\end{figure}

\subsection{Segmentation of WM based on multi-subject atlas}
The method proposed in \cite{labra2017fast} classifies WM subject fibers based on a multi-subject bundle atlas \cite{Guevara12}. Subject fibers are classified in an atlas bundle by computing the maximum Euclidean distance between each subject fiber and each atlas centroid, and keeping only fibers with a distance under a user-specified threshold defined for each bundle. The maximum Euclidean distance ($d_{ME}$) between an N-point fiber $a$ and centroid $b$, is based on the Euclidean distance $d_{E}$ between its corresponding 3D points $a_i$ and $b_i$, where:

\begingroup 
\setlength{\abovedisplayskip}{-10pt}
\setlength{\belowdisplayskip}{2pt}
\begin{equation}
d_E(a_i,b_i) = \norm[2]{(a_i - b_i)}
\end{equation}

\begin{equation}
\label{eq:dME}
d_{ME}(a,b) = \min (\max_i(d_E(a_i,b_i)), \max_i(d_E(a_i,b_{N-i})))
\end{equation}


\endgroup 

Because the spatial orientation of fibers on the dataset is unknown, $d_{ME}$ considers the minimum of both possible orientations (direct and inverse), as shown in equation \ref{eq:dME}.
The $d_{ME}$ distance is then normalized by adding a term $\textit{TN}$, which penalizes the difference between the lengths of the subject fiber ($l_S$) and atlas centroid ($l_C$):
\vspace{-8pt}

\begingroup 
\setlength{\abovedisplayskip}{-10pt}
\setlength{\belowdisplayskip}{1pt}
\begin{equation}
\textit{TN} = {\left( \frac{\abs{(l_S-l_C)}}{\max(l_S, l_C)} + 1 \right)}^2 - 1
\end{equation}
\endgroup 

Before processing, subject fibers are resampled using 21 equidistant points. The algorithm classifies subject fibers in two steps. First, the \textit{pre-classification} stage discards fibers by only using a subset of 3D points to compute $d_{ME}$, using the fact that if $d_E(a_i,b_i)$ is above the threshold for just one value of $i$, then $d_{ME}(a,b)$ will be above the threshold, and fiber $a$ will be discarded. Next, the \textit{classification} step computes the complete distance metric ($d_{ME} + \textit{TN}$) for the fibers that were not discarded in the first stage. 


The optimization of the execution time of the algorithm is mainly achieved through:

\textbf{Progressive fiber discarding}: for fibers resampled with 21 points, it is necessary to calculate 42 distances $d_E(a_i,b_i)$, considering direct and inverse directions. 
To reduce the number of calculations, the algorithm discards fibers with the simplified distance metric using first only the center point (\textit{test1}). Then, the surviving fibers are discarded using the end points (\textit{test2}), and the third stage (\textit{test3}) discards surviving fibers using 4 intermediate points. Finally, the remaining fibers are classified using the complete distance metric (\textit{test4}).


\textbf{Parallel execution}: the dataset is divided into smaller subsets to fit in the computer memory.
For each subset, the distances between subject fibers and atlas centroids are evaluated in parallel using multiple threads. The subsets are processed sequentially.


\begin{algorithm}[ht]
\caption{Parallel segmentation}
\label{alg:parallel}
\begin{algorithmic}[1]

\FOR {i = 0 \textbf{to} nfibersSubject}
	\FOR {j = 0 \textbf{to} atlasData.size()}
		\FOR {k = 0 \textbf{to} (atlasData[j].size()/ndataFiber)}
			\STATE{$isInverted, isDiscarded \leftarrow false$}
    		\STATE{$ed \leftarrow $-$1$}
    		\STATE{$isDiscarded $ = $ discardCenter()$} \COMMENT{test1}
    		\STATE{\textbf{if} isDiscarded \textbf{then} continue \textbf{end if}}
    		\STATE{$isDiscarded $ = $ discardExtremes()$} \COMMENT{test2}
    		\STATE{\textbf{if} isDiscarded \textbf{then} continue \textbf{end if}}
    		\STATE{$isDiscarded $ = $ discardFourPoints()$} \COMMENT{test3}
    		\STATE{\textbf{if} isDiscarded \textbf{then} continue \textbf{end if}}
    		\STATE{$isDiscarded $ = $ discarded21points()$} \COMMENT{test4}
    		\IF{$ed \neq $-$1$} 
    			\IF{$ed < euclideanDistances[i]$}
    				\STATE{$euclideanDistances[i] \leftarrow ed$} \COMMENT{gets minimum euclidean distance}
    				\STATE{$assignment[i] \leftarrow j$} \COMMENT{each fiber is assigned to the corresponding bundle}
    			\ENDIF
    		\ENDIF
    	\ENDFOR
	\ENDFOR
\ENDFOR
\RETURN ${assignment}$ \COMMENT{returns the bundle associated with each fiber}
\end{algorithmic}
\end{algorithm}

\vspace{-8pt}
\subsection{New optimized algorithm}
We propose Algorithm \ref{alg:parallel} to optimize both memory and computation time. 
The strategy of using the 4 stages to progressively discard fibers has been maintained, but several optimizations have been included in the discarding process. Next, we describe the most relevant optimizations within the code.

\textbf{Amount of reserved memory}: the space complexity of the original algorithm is $\mathcal{O}(NM)$, where $N$ is the number of fibers in the subject dataset, and $M$ is the number of centroids in the atlas. For large sets, this requirement is unfeasible and the data must be processed sequentially in chunks, which limits the parallelism of the algorithm. In contrast, the space complexity of the new algorithm is only $\mathcal{O}(N+M)$, because it compares one subject fiber against each atlas centroid until a label is assigned to it. This enables us to load the entire dataset into memory and process all fibers in parallel, even if the computer has limited memory or swap file resources.


  
\textbf{Parallel synchronization}: intuitively, the algorithm discards fibers in  parallel independently of each other, so it has no collision problems. This is achieved using an array of size equal to the number of subject fibers, where the index of the bundle to which it belongs is stored. 
  
\textbf{Data locality}: the proposed parallel algorithm improves the locality of the memory accesses, thus making more efficient use of the local memory system of each processor. This is possible due to the nesting of the loops, in which each processor can keep the candidate fibers for each bundle in their cache and RAM.

\textbf{Inference of fiber direction}: according to equation \ref{eq:dME}, the original algorithm calculates the distance between fibers and centroids in both directions in each stage, which leads to duplicated operations. Our algorithm uses the distances computed in second discarding stage to determine the direction of the fiber. Therefore, the third and fourth stage are evaluated in only one direction, thus halving the amount of computation and eliminating the $\min()$ operation of equation \ref{eq:dME} .

Another improvement is that the algorithm evaluates the discarding criterion after each pairwise distance calculation $d_{E}(a_i,b_i)$. If the distance is above the threshold for the corresponding bundle, the evaluation is stopped and the fiber is immediately discarded. 

Finally, the final classification stage of the previous version adds the normalization term $\textit{TN}$ to all the distances before comparing them to the threshold. In our new algorithm, if $d_{ME}$ is larger than the threshold, it discards the fiber immediately without computing $\textit{TN}$.
 
\textbf{Classification of fibers}: The original algorithm, tested with deep WM (DWM) bundles, can assign a fiber to several bundles. This is not a problem with the DWM atlas \cite{Guevara12}, because it rarely labels a fiber with two bundles. With its original parallel implementation, it was not possible to efficiently update the minimum distance for several bundles and keep the minimum. For SWM bundles, is very common to satisfy the threshold criterion for two neighboring bundles, so we modified the algorithm to improve the classification and perform the labelling for only the closest bundle.



\section{Results}
\label{sec:res}
The optimized algorithm was implemented in C++ 11, compiled with gcc 7.3.0 (option -O3). We ran our experiments on a computer with an 8-core Intel Core i7-6700K CPU running at 4GHz, 8MB of shared L3 cache and 8GB of RAM, using Ubuntu 18.04.1 LTS with kernel 4.15.0-36 (64 bits).

\begin{figure}[t!]
	
	\begin{minipage}[b]{1.0\linewidth}
		\centerline{\includegraphics[width=8.5cm]{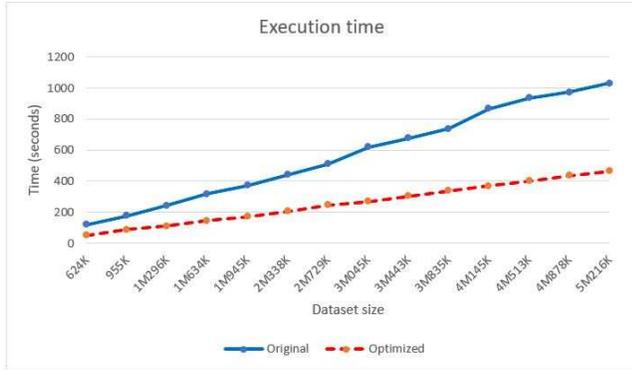}}
	\end{minipage}
	\caption{Execution time of both algorithms versus subject fiber dataset size.}
	\label{fig:time}
	
\end{figure}

To evaluate the performance of the algorithms, we used a deterministic tractography dataset of one subject. The optimized version of the algorithm produces almost the same results (Figure \ref{fig:atlasSubject}.B) as the previous algorithm, except that it classifies fibers to only their closest bundle below the distance threshold. The experiments use resampled subject datasets from 600,000 to about 5,216,000 fibers.  
Figure \ref{fig:time} shows the execution time of both algorithms for different subject dataset sizes, using the atlas of 62 bundles \cite{roman2017clustering}.
For a dataset of 1,634,000 fibers, our new parallel algorithm runs 2.15 times faster than its previous version. With a dataset of 4,145,000 fibers, the algorithm is 2.34 times faster than its predecessor. The original algorithm segments a subject dataset of 5,216,000 fibers in approximately 17 min, while our algorithm segments the same subject in less than 8 min. 

\begin{figure}[t!]
	
	\begin{minipage}[b]{1.0\linewidth}
		\centerline{\includegraphics[width=8.5cm]{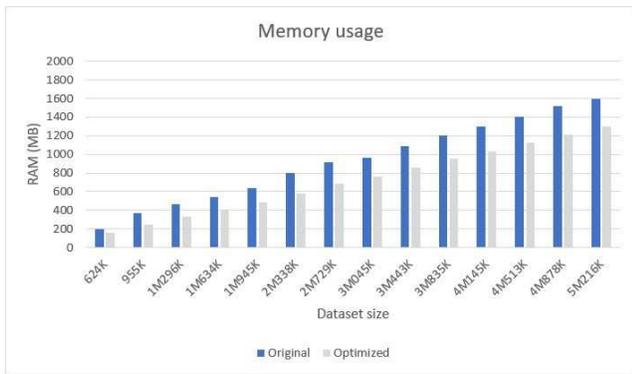}}
		\caption{Memory usage for both algorithms versus subject fiber dataset size.}
		\label{fig:mem}
	\end{minipage}
\end{figure}

Figure \ref{fig:mem} shows the memory used by both algorithms using the atlas composed of 100 bundles \cite{guevara2017reproducibility}. For the first algorithm, the memory usage reflects the division of the subject fiber set in chunks that the algorithm performs in order to fit the data in memory. In our new algorithm, the graph shows the memory reserved by the algorithm for the entire dataset. Even with the complete dataset in memory, the new algorithm consistently consumes less memory than the original. As an example, for the dataset of 3,443,000 fibers, the previous algorithm consumes 1.27 times more RAM than the new one. Memory usage grows linearly for both algorithms because the atlas size is fixed.

The source code of the new implementation, along with its tutorial, are available from the corresponding author upon request, to \textit{pamela.guevara@gmail.com}.

\section{Conclusions}
\label{sec:con}
We have presented a new parallel algorithm for the segmentation of WM fibers. The optimized algorithm improves upon both the execution time and memory usage of its predecessor. Using a subject dataset of 4,145,000 fibers with an atlas of 7,753 fibers \cite{guevara2017reproducibility} on a computer with an Intel i7-6700K processor and 32 GB of RAM, the new algorithm executes in 6 minutes, a speedup of 2.34 times over the previous version. In addition, for the same number of fibers, it reduces memory usage by a factor of 0.79. The new algorithm has linear space complexity, which allows it to load the entire dataset and exploit more parallelism than the original version, even on computers with limited memory and swap space. Also, the new algorithm is more scalable, opening the possibility of applying it to more massive tractography datasets and bigger atlases, in particular, those obtained from superficial white matter fibers, based on probabilistic tractography, which can reach several million of fibers. Furthermore, the new algorithm performs a better classification, labelling the fibers with only the closest bundle. Also, the new implementation in C++ enables easier future extensions and maintenance of the code.



\bibliographystyle{IEEEbib}
{\small \bibliography{strings,refs}}

\end{document}